\pdfoutput =  1

\documentclass[%
preprint,%
 amssymb, amsmath,%
 prb,%
 superscriptaddress,
]{revtex4-1}

\usepackage{bm}%
\usepackage[colorlinks=true,linkcolor=blue]{hyperref}%
\expandafter\ifx\csname package@font\endcsname\relax\else
 \expandafter\expandafter
 \expandafter\usepackage
 \expandafter\expandafter
 \expandafter{\csname package@font\endcsname}%
\fi
\usepackage{graphicx}
\usepackage[utf8]{inputenc}   
\usepackage{dcolumn}
\usepackage{color}
\usepackage{xcolor}
\usepackage{bm}
\usepackage{hyperref}
\usepackage{amsmath}
\begin{document}

\preprint{APS/123-QED}

\title{The role of electron and phonon temperatures in the helicity-independent all-optical switching of GdFeCo}

\author{J. Gorchon}%
\email{jgorchon@lbl.gov}
\affiliation{Lawrence Berkeley National Laboratory, 1 Cyclotron Road, Berkeley, CA 94720, USA}
\affiliation{Department of Electrical Engineering and Computer Sciences, 
University of California, Berkeley, CA 94720, USA}

\author{R. B. Wilson}%
\email{rwilson@engr.ucr.edu}
\affiliation{Lawrence Berkeley National Laboratory, 1 Cyclotron Road, Berkeley, CA 94720, USA}
\affiliation{Department of Electrical Engineering and Computer Sciences, 
University of California, Berkeley, CA 94720, USA}
\affiliation{Department of Mechanical Engineering and Materials Science \& Engineering Program, University of California, Riverside, CA 92521, USA}

\author{Y. Yang}
\affiliation{Department of Materials Science and Engineering, 
University of California, Berkeley, CA 94720, USA}

\author{A. Pattabi}%
\affiliation{Department of Electrical Engineering and Computer Sciences, 
University of California, Berkeley, CA 94720, USA}

\author{J. Y. Chen}
\affiliation{Department of Electrical and Computer Engineering, University of Minnesota, Minneapolis, MN 55455, USA}

\author{L. He}
\affiliation{Department of Electrical and Computer Engineering, University of Minnesota, Minneapolis, MN 55455, USA}

\author{J.P. Wang}
\affiliation{Department of Electrical and Computer Engineering, University of Minnesota, Minneapolis, MN 55455, USA}

\author{M. Li}
\affiliation{Department of Electrical and Computer Engineering, University of Minnesota, Minneapolis, MN 55455, USA}

\author{J. Bokor}
\affiliation{Lawrence Berkeley National Laboratory, 1 Cyclotron Road, Berkeley, CA 94720, USA}
\affiliation{Department of Electrical Engineering and Computer Sciences, 
University of California, Berkeley, CA 94720, USA}

\date{\today}

\begin{abstract}
Ultrafast optical heating of the electrons in ferrimagnetic metals can result in all-optical switching (AOS) of the magnetization. Here we report quantitative measurements of the temperature rise of GdFeCo thin films during helicity-independent AOS. Critical switching fluences are obtained as a function of the initial temperature of the sample and for laser pulse durations from $55$ fs to $15$ ps. We conclude that non-equilibrium phenomena are necessary for helicity-independent AOS, although the peak electron temperature does not play a critical role. Pump-probe time-resolved experiments show that the switching time increases as the pulse duration increases, with $10$ ps pulses resulting in switching times of $\sim 13$ ps. These results raise new questions about the fundamental mechanism of helicity-independent AOS.
\end{abstract}

\maketitle



Ultrafast optical excitation of magnetic materials causes distinctive dynamics of great interest for applications~\cite{Thiele2002,He2015,LeGuyader2012} and fundamental science~\cite{Beaurepaire1996,Radu2011,Kirilyuk2013}. For example, short-pulse laser irradiation of a magnetic thin film can reverse the direction of the magnetic moment, even in the absence of an external magnetic field, a phenomena known as all optical switching (AOS)~\cite{Stanciu2007,Radu2011, Ostler2012a,Mangin2014,Lambert2014b}. Many AOS studies have only observed deterministic switching if the laser pulse irradiating the sample is circularly polarized~\cite{Stanciu2007,Mangin2014,Lambert2014b}. However, in ferrimagnetic GdFeCo films, AOS is observed with linear polarized light and has been described as an ultrafast thermal effect~\cite{Radu2011, Ostler2012a}.

Despite intense study, the mechanisms of AOS remain unclear due to the rich physics that are found after a sub-100 femtosecond pulsed laser excitation. In the first hundred femtoseconds, highly non-equilibrium phenomena such as non-thermal carrier excitation~\cite{Guidoni2002,Mueller2013} and super-diffusive spin-currents~\cite{Battiato2010} may take place. In the next few hundred femtoseconds, electrons become thermalized with each other resulting in a high electronic temperature $T_e$, but remain out of thermal equilibrium with the lattice and spin degrees of freedom~\cite{Beaurepaire1996}. In addition to these nonequilibrium phenomena, the strong dependence of equilibrium magnetic properties on temperature could also play a central role in AOS~\cite{Thiele2002,Stanciu2006}, as it does in heat assisted magnetic recording (HAMR) technology~\cite{Thiele2002}. Finally, magneto-optical phenomena such as the inverse Faraday effect~\cite{Kimel2004b} or magnetic circular dichroism (MCD)~\cite{Khorsand2012} complete the wide range of coexisting mechanisms that may play a role in AOS, making it a fascinating but challenging problem to understand.

The energy absorbed by the metal film, and the resulting transient temperature response, are known to play a central role in ultrafast demagnetization of single element ferromagnets ~\cite{Beaurepaire1996, Roth2012, Kimling2014}. However, due to the large number of mechanisms that may contribute to AOS, it has been difficult to determine the primary role of energy and temperature during AOS. Temperature rise can directly or indirectly facilitate magnetization switching in a number of ways. For example, in HAMR, the lattice temperature $T_p$ of the system is heated close to the critical Curie temperature $T_C$ to reduce the anisotropy before an applied field favors a particular direction for the magnetization upon cooling~\cite{Thiele2002}. In contrast, AOS models for ferrimagnets~\cite{Radu2011, Ostler2012a, Mentink2012, Barker2013, Wienholdt2013,Atxitia2015,Kalashnikova2016,Schellekens2013b} do not require the lattice temperature of the film to approach the Curie temperature.  Instead, these models rely on transient electron temperatures that are out of equilibrium with the lattice to induce the dynamics of the Gd and Fe magnetic sublattices~\cite{Radu2011}.




Despite the centrality of temperature to prevailing theories for AOS, the energy required for switching, and the resulting temperature response of the electrons and phonons remains unclear. This is largely related to uncertainties in the minimum absorbed fluence required for switching (i.e. critical fluence $F_C$) and unknown thermal parameters. Peak temperatures and subsequent cooling are determined by $F_C$, the electron phonon coupling parameter $g_{ep}$, and the electronic heat capacity $C_e$.  $C_e$ and $g_{ep}$ are generally set by assuming \emph{typical} values for transition metals. However, reported values for $g_{ep}$ for transition metals vary by an order of magnitude~\cite{Bonn2000,Atxitia2015}. 

Indeed, reported $F_C$ values for GdFeCo vary from 0.75 mJ/cm$^2$~\cite{Vahaplar2012} to 3.14 mJ/cm$^2$~\cite{Vahaplar2012}. As an example, assuming the carefully determined threshold from Ref.~\cite{Khorsand2012} $F_C=2.6\pm0.2$ for a $d=20$ nm thick film and a total heat capacity of $C=3\pm 0.2$ MJ/(m$^3$K)~\cite{Hellman1998}, the transient $T_p$ can be calculated through $T_p=T_0+F_C/(d*C)$, 
where $T_0$ is the initial temperature. $T_p$ should rise to about $750$ K, well above $T_C\approx550$ K~\cite{Vahaplar2012}. Crossing $T_C$ would imply a loss of memory of the magnetic order. The final magnetization would then be determined by the cooling conditions, analogous to HAMR, which is in contrast with what most AOS models assume~\cite{Radu2011, Ostler2012a, Mentink2012, Barker2013, Wienholdt2013,Atxitia2015,Kalashnikova2016,Schellekens2013b}. 


In this work, we carefully measure $F_C$ for the helicity-independent AOS of GdFeCo films, through single shot switching and stroboscopic pump-probe experiments. $F_C$ values are then obtained as a function of the sample temperature $T_0$ and the laser pulse duration $\Delta t$. We observe AOS for pulse durations as long as $\Delta t=15$ ps and identify two distinct mechanisms that prevent AOS at longer pulse durations. By using the three temperature model~\cite{Kimling2014}, we calculate that for $\Delta t=55$ fs, $T_e$ reaches $\sim 1600$ K, while for a $\Delta t=12.5$ ps pulse $T_e$ reaches $\sim 530$ K. We conclude that the electron peak temperature does not play a key role in the switching mechanism, and raise questions about the conclusions in various AOS models. Finally, we performed pump-probe experiments as a function of the pulse duration and showed that $10$ ps pulses result in switching times of $\sim 13$ ps.

\begin{figure}
\includegraphics[trim=0cm 0cm 0cm 0cm, clip=true, width=0.9\columnwidth]{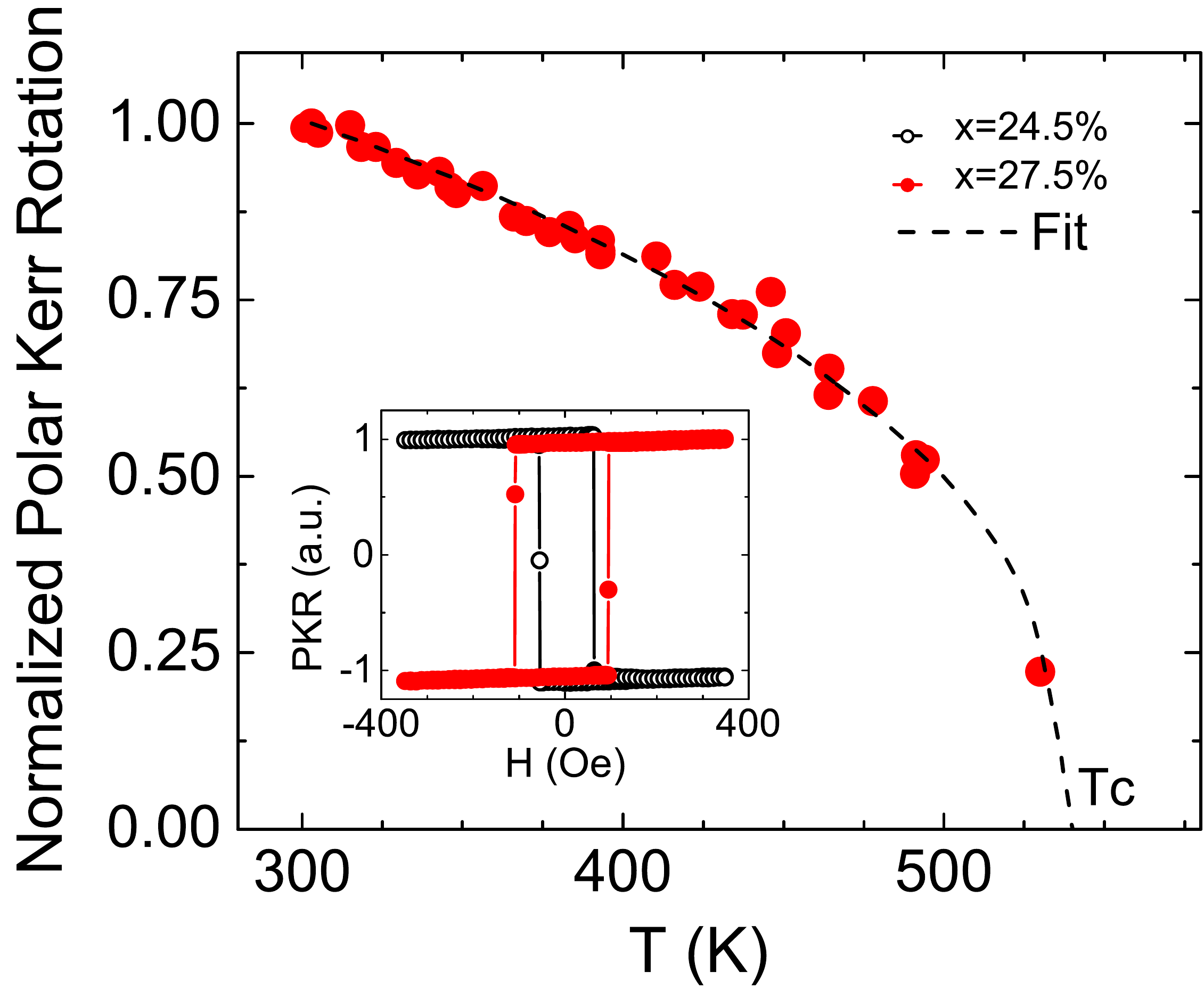}
\caption{\label{fig:MOKE}: Temperature dependence of the normalized Kerr rotation of Gd$_x$(Fe$_{90}$Co$_{10}$)$_{100-x}$ with $x=27.5$\%. As inset, the magnetic hysteresis as a function of the out-of-plane external field $H$, at room temperature, for samples with $x=27.5$\% and $x=24.5$\%.
}
\end{figure}

The experiments were carried out on two Gd$_x$(Fe$_{90}$Co$_{10}$)$_{100-x}$ films of concentrations $x=24.5$\% and $27.5$\% grown by co-sputtering of the following stacks (in nm): Si/SiO$_2$($300$)/Ta($2.5$)/GdFeCo($14$)/Ta($3.6$) /Ta$_2$O$_5$($2.8$). The layer thicknesses were confirmed by X-ray reflectivity. Both samples exhibited perpendicular magnetic anisotropy, which was determined via magneto-optic Kerr effect (MOKE) hysteresis measurements. A Curie temperature of about $540$ K was obtained by fitting the normalized polar Kerr rotation (NPKR) via phenomenological formula~\cite{Dunlop97} NPKR $= [(T-T_C)/(T-300)]^{0.39} $ (see Fig.~\ref{fig:MOKE}). This Curie temperature is close to previously reported values~\cite{Kirilyuk2013}. The compensation temperature $T_M$ was measured by monitoring the coercivity and polarity of the magnetic hysteresis via MOKE as the sample was heated with an electric heater~\cite{Kirilyuk2013}. We found $T_M\approx 430K$ for sample $x=27.5$\%. Sample $x=24.5$\% presented a hysteresis with the opposite polarity to that of  $x=27.5$\% at room temperature (see inset of Fig.~\ref{fig:MOKE}), meaning its compensation temperature was below room temperature. We did not have the capability to measure below ambient temperature. The two samples will respectively be addressed as Gd$_{24}$FeCo and Gd$_{27}$FeCo throughout the text.

An amplified $250$ kHz Ti:sapphire laser with $810$ nm center wavelength was used for generating the high energy pulses and as a time-resolved probe (Coherent RegA). The laser pulse duration full-width at half maximum (FWHM) was tunable from $\Delta t=55$ fs to $\Delta t=25$ ps by adjusting the final pulse compressor in the chirped pulse amplifier~\cite{Pessot1987}.  Individual single-shot laser pulses could be obtained from our laser system. A MOKE microscope was used for imaging the sample magnetization after each single laser pulse shot.

\begin{figure}
\includegraphics[trim=0cm 0cm 0cm 0cm, clip=true, width=1.1\columnwidth]{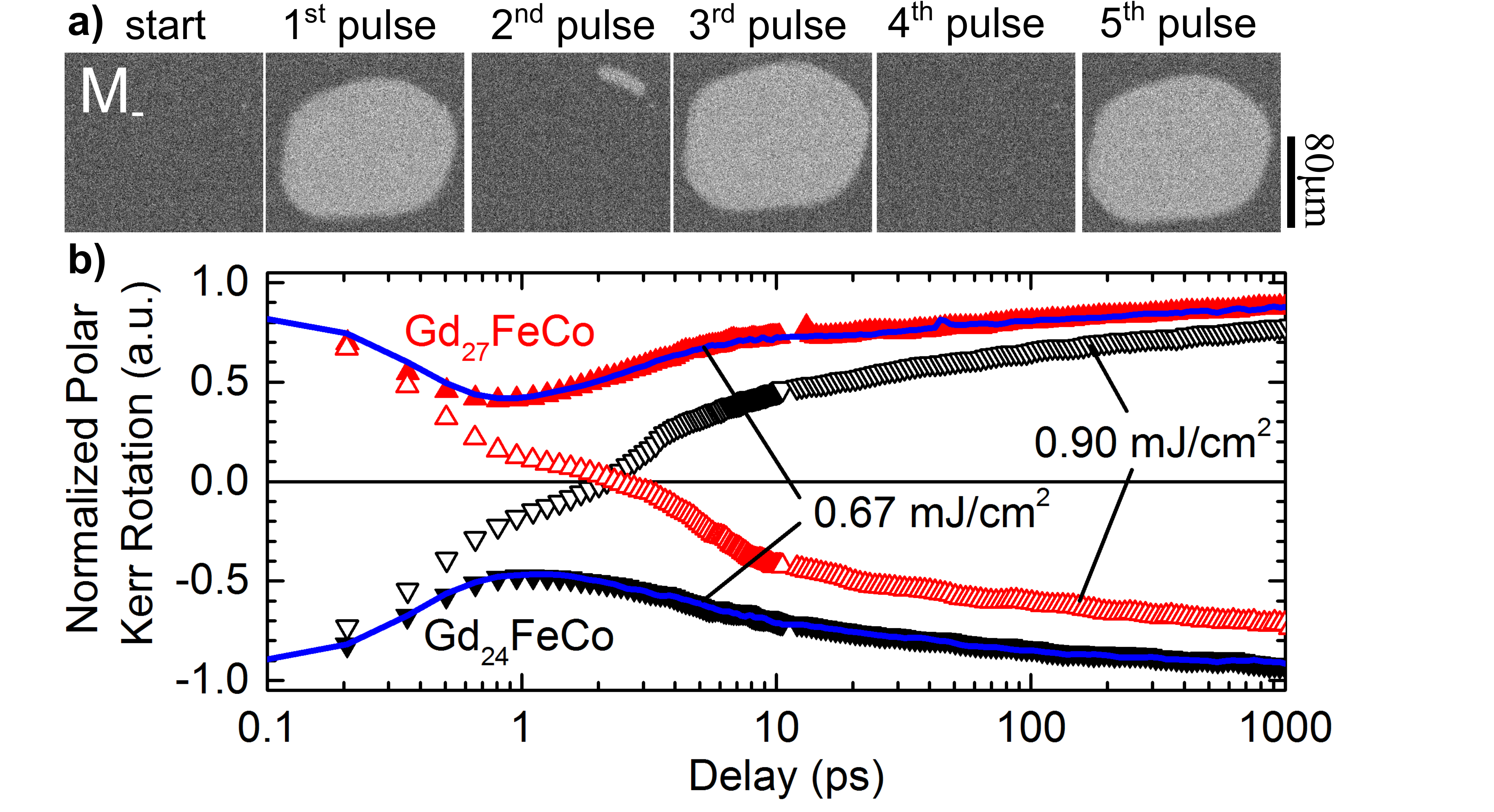}
\caption{\label{fig:AOS_toggle}: a) MOKE micrographs of the magnetization of the Gd$_{27}$FeCo film exposed to successive linearly polarized laser pulses of $\Delta t=55$ fs on an initially 'down' ($M-$) magnetized sample. Reliable all-optical switching of the magnetization independently of the helicity of light is demonstrated. b) Evolution of the normalized polar Kerr rotation of GdFeCo samples induced by a linearly polarized $\Delta t=55$ fs pump, under a constant perpendicular external field of $55$ Oe. The blue lines correspond to the evolution under no external field, which show no difference during the first nanosecond.
}
\end{figure}

We first discuss single shot experiments. In these experiments, the laser beam was incident with an angle of $40^{\circ}$ with respect to the normal. The spatial beam profile was obtained by the knife-edge technique\cite{Magnes2006a} and the energy of each pulse was monitored with a calibrated photodiode connected to a 6 GHz oscilloscope. To accurately determine the fluence absorbed in the GdFeCo film, a multilayer absorption calculation was performed~\cite{Hecht2002} using an effective index of refraction of $n=3.7+4.2i$ for Ta/GdFeCo/Ta measured by ellipsometry. An absorption of $29$\% was found (see the absorption profile in the supplementary materials~\cite{SuppMat}). The magnetization of the film was saturated with an external magnetic field $H\approx\pm100$ Oe. Following removal of the external field, the film was then exposed to a single linearly polarized laser pulse. As shown in Fig.~\ref{fig:AOS_toggle}, after each laser pulse of the same energy, the magnetization in a small region reliably toggles between white ('up') and black ('down') back and forth. Our observation of helicity-independent toggling of the GdFeCo magnetization is consistent with the helicity independent AOS reported in Ref.~\cite{Ostler2012a}. 

\begin{figure*}
\includegraphics[trim=0cm 0cm 0cm 0cm, clip=true, width=1\columnwidth]{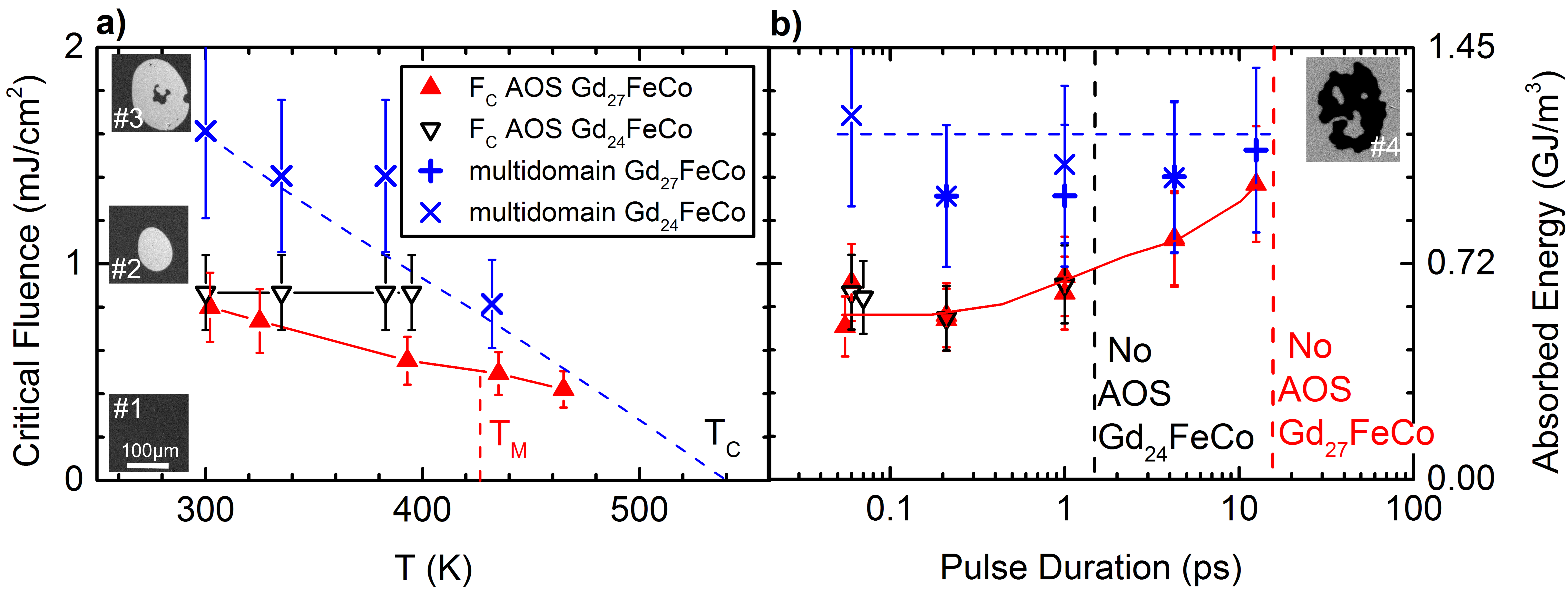}
\caption{\label{fig:Fcrit_vs_singleShot}: Critical fluence $F_C$ for AOS and multidomain states as a function of (a) the initial temperature of the sample, for $\Delta t=55$ fs laser pulses and (b) the laser pulse duration at room temperature. Solid lines are guides for the eyes. The blue dashed lines are a calculation of the fluence needed to make the lattice reach $T_C$ (see text). MOKE images in (a) show the typical result in each fluence range. From bottom to top: No switch (ultrafast demagnetization), AOS and multidomain state. The vertical dashed lines in (b) show the limits for observation of AOS in each sample. The right hand image shows the fully demagnetized state obtained for a $\Delta=16$ ps pulse of $\sim 1.85$ mJ/cm$^2$ on Sample Gd$_{27}$FeCo.}
\end{figure*}

In the absence of domain wall motion, the reversed domain size is determined by the area within the Gaussian laser profile with a fluence above $F_C$~\cite{Khorsand2012}. However, in our films, we noticed that domain wall motion reduces the size of the reversed domain in the seconds following laser irradiation. We observe a critical domain size ($\approx 10$ $\mu$m) below which optically switched domains shrink and collapse after several seconds. Instability of small magnetic domains is a well understood phenomena that occurs whenever the domain wall energy is larger than the domain stabilizing pinning and dipolar energy terms~\cite{Malozemoff1979}. In order to minimize the effect of such relaxation mechanisms on our measurement of the critical fluence, the pump diameter (FWHM) was chosen to be relatively large ($\approx 0.16$ mm). The absorbed critical fluences $F_C$ shown in Fig.~\ref{fig:Fcrit_vs_singleShot} were then obtained by decreasing the pump fluence until no switching was observed. For $T=300$ K and $\Delta t=55$ fs, we found $F_C=0.82\pm0.16$ mJ/cm$^2$.

We performed time-resolved pump-probe MOKE measurements on both samples. For these experiments, a constant, perpendicular external field of $55$ Oe was applied to reset the magnetization between pump pulses. The pump beam, incident at $40^{\circ}$, had a spot diameter (FWHM) of $\sim 100$ $\mu$m, whereas the probe, at normal incidence, was kept much smaller with a spot diameter of $\sim 6$ $\mu$m. As shown in Fig.~\ref{fig:AOS_toggle}.b for a fluence of $0.86$ mJ/cm$^2$ the reversal occurs, against the external magnetic field, for both samples within a few picoseconds. The opposite sign of the signal at negative time delay for samples with $T_M$ above and below room temperature is due to the sensitivity of our $810$ nm probe to the FeCo sublattice magnetization~\cite{Stanciu2007a}. When $T<T_M$ the external field drives the dominant Gd sublattice whereas at $T>T_M$ the field drives the FeCo~\cite{Stanciu2007a}.

\begin{figure}
\includegraphics[trim=1cm 0.5cm 2cm 0cm, clip=true, width=1\columnwidth]{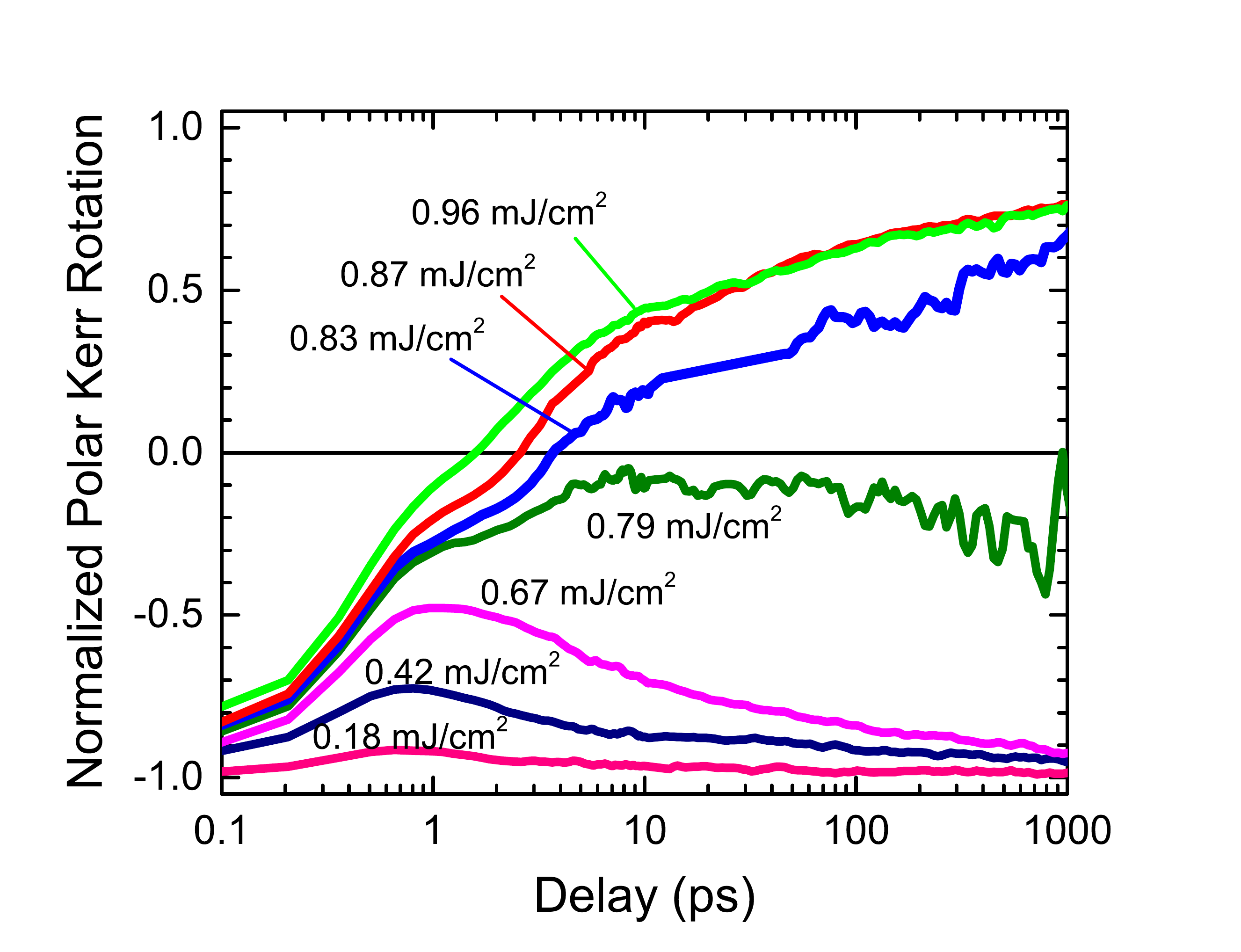}
\caption{\label{fig:pumpprobe}: In solid lines, the evolution of the magnetization of Gd$_{24}$FeCo after a $\Delta t=55$ fs linearly polarized pump pulse, at room temperature. The switching threshold is close to $0.8$ mJ/cm$^2$, agreeing with the single shot experiments (see Fig.~\ref{fig:Fcrit_vs_singleShot}).}
\end{figure}

For comparison of $F_C$ obtained through pump-probe experiments with $F_C$ from single shot experiments, we first checked that the switching behavior was not affected by the constant external field. For this purpose, pump-probe demagnetization experiments at low fluences with no external field were performed on both samples (blue lines in Fig.\ref{fig:AOS_toggle}.b). No difference in the Kerr signal was observed with respect to experiments performed with the $55$ Oe external field. In addition, we performed experiments on sample Gd$_{24}$FeCo (black down triangles in Fig.\ref{fig:AOS_toggle}.b) at $T>T_M$ where no transition through $T_M$ was possible due to laser heating. This means that field induced switching scenarios due to crossing of $T_M$ can be discarded~\cite{Stanciu2007a}. Fig.\ref{fig:pumpprobe} shows the fluence dependence of the magnetization evolution in Gd$_{24}$FeCo at $T=300$ K and for $\Delta t=55$ fs. The curve at $0.79$ mJ/cm$^2$ presents relatively higher noise at long time delays, which we interpret as the result from laser intensity fluctuations when the fluence approaches $F_C$. We thus find $F_C\approx0.8$ mJ/cm$^2$, which is consistent with our single shot technique for measuring critical fluences (see Fig.~\ref{fig:Fcrit_vs_singleShot}).

In order to study the importance of the lattice temperature in AOS, the critical fluence was recorded as a function of the initial temperature $T_0$, which was varied by mounting the sample on a resistive heater. A threshold at which a multidomain pattern was observed was also recorded (see pic \#3 in Fig.~\ref{fig:Fcrit_vs_singleShot}.a). The measurement of this threshold has a large uncertainty due to the stochastic nature of domain nucleation and the instability of the small multidomain patterns. Within experimental accuracy we found the multidomain thresholds for both samples to be equal (see  Fig.~\ref{fig:Fcrit_vs_singleShot}.b).

In the case where the whole system reaches $T_C$, a multidomain magnetization pattern is expected to arise as the sample cools down from the paramagnetic state and is remagnetized randomly. Indeed, the transition from pure AOS to multidomain is observed (pics \#2 and \#3 in Fig.~\ref{fig:Fcrit_vs_singleShot}) at a particular threshold fluence (blue crosses in Fig.~\ref{fig:Fcrit_vs_singleShot}). Transient temperatures for electrons and the lattice were calculated with the three temperature model~\cite{Kimling2014}, and the threshold at which the overall temperature (back in equilibrium) exceeds $T_C$ is plotted as a blue dashed line in Fig.~\ref{fig:Fcrit_vs_singleShot}. The model will be discussed later in the text. As both samples have very similar compositions, resulting in similar total heat capacities and Curie temperatures, the transient equilibrium temperature and thus the multidomain fluence threshold are expected to be similar. Therefore, the demagnetization threshold sets a limit above which no AOS can be observed.

The critical fluence for Gd$_{24}$FeCo is independent of ambient temperature, while the critical fluence of Gd$_{27}$FeCo decreases by a factor of two upon a change in ambient temperature from $300$ to $450$ K (Fig.~\ref{fig:Fcrit_vs_singleShot}). We believe the different temperature dependence is related to the difference in energy transfer rates between sublattices in both samples, as has been predicted~\cite{Barker2013}. A discussion on the energy transfer rates will follow later in the text.  However, both samples display a weaker temperature dependence than we would expect if AOS was an equilibrium phenomena analogous to HAMR. If changes to equilibrium magnetic properties were the primary driver of AOS in a manner analogous to HAMR, the peak lattice temperature reached following laser irradiation at $F_C$ would be insensitive to ambient temperature. At ambient temperatures of $300$ K and $470$ K, the calculated transient temperature rise in the lattice following irradiation is $\sim 150$ K and $\sim 70$ K, respectively. Therefore, the peak lattice temperature during AOS varies from $450$ K to $540$ K ($\sim T_C$) for ambient temperatures from $300$ to $470$ K. Therefore, as expected, we confirm that unlike in HAMR, heat induced changes to equilibrium magnetic properties are not the primary driver of AOS.

As discussed above, numerous models predict that the transient temperature response of the electrons $T_e$ following laser irradiation is responsible for AOS~\cite{Radu2011, Ostler2012a, Mentink2012, Barker2013, Wienholdt2013,Atxitia2015}. In atomistic calculations typically $T_e$ is coupled to a Langevin random field term which is then entered into a Landau-Lifshitz-Gilbert equation. The equations are,

\begin{eqnarray}
\label{eq:LLG}
\frac{\delta S_i}{\delta t}=-\frac{\gamma_i}{(1+\alpha_i^2)\mu_i} \left(S_i \times H_i+\alpha_i S_i \times \left[S_i \times H_i \right] \right)
\end{eqnarray}

where $S_i$ is the reduced atomic localized spin, $\gamma_i$ is the gyromagnetic ratio of the sublattice, $\alpha_i$ is an effective damping parameter (a channel to dissipate angular momentum to the lattice) and $H_i$ is an effective field given by,

\begin{eqnarray}
\label{eq:Hi}
H_i=\eta_i + \frac{\delta E_i}{\delta S_i}
\end{eqnarray}

where $\eta_i\propto \alpha_i T_e$ is the Langevin noise of the sublattice, proportional to $T_e$, and $E_i$ is the energy of the sublattice, including exchange, anisotropy and Zeeman terms. These models can successfully reproduce the switching behavior through a three step process. In the first step $T_e$ has to quickly overcome $T_C$ in order for the Langevin field (thermal excitations) to overcome the exchange field. This induces the independent demagnetization of the sublattices. Due to their different damping (rate of dissipation of angular momentum) and magnetic moments, demagnetization for different lattices will occur at different rates, the Fe demagnetizing faster~\cite{Radu2011}. The second step involves the cooling of $T_e$ which allows the remagnetization of the completely demagnetized Fe sublattice. At this stage the exchange fields become dominant again, and as the Gd demagnetizes towards its equilibrium magnetization (at $T_e$) conservation of angular momentum induces the switching of the Fe sublattice. The third step consists in the antiparallel alignment of the Gd spins relative to the Fe spins due to the exchange interaction. However, in these models it is often clearly claimed~\cite{Radu2011, Ostler2012a, Mentink2012} that initially $T_e$ needs to quickly overcome $T_C$ in order to decouple the sublatices and allow a faster demagnetization of the Fe sublattice. Other microscopic models~\cite{Kalashnikova2016,Schellekens2013b} that treat the energy and angular momentum exchange through scattering processes, also reach similar conclusions, and state the necessity of short and intense pulses for the initial demagnetization of both sublattices to happen at different rates.

To test the importance of the peak electron temperature, single shot AOS experiments as a function of the pulse duration (FWHM) $\Delta t$ were performed. As $\Delta t$ increases, the laser peak intensity drops as $1/\Delta t$ resulting in a lower peak $T_e$. However, since energy transfer rates depend on temperature differences between heat baths, electrons actually lose less energy when they are cool. The result is a drop of the peak $T_e$ by a factor of $\sim3$ when going from $\Delta t=50$ fs to $10$ ps~\cite{Kimling2014}. If the peak $T_e$ is a key parameter for AOS, as $\Delta t$ is increased the critical fluence should increase proportionally. We observe a relatively weak dependence of the critical fluence on the pulse duration (see Fig.\ref{fig:Fcrit_vs_singleShot}.b). The energy needed for AOS increases $\sim 50$\% as the pulse duration increases by over two orders of magnitude. Similar trends have been reported in the context of helicity-dependent AOS~\cite{Vahaplar2012,Steil2011a}. However the analysis in Refs.~\cite{Vahaplar2012,Steil2011a} was made in terms of helicity-induced opto-magnetic fields. Furthermore, high critical fluences were reported that would easily heat the lattice above $T_C$. As we have shown, such high lattice transient temperatures should result in a random multidomain state instead of a helicity-independent AOS. In this work, as shown in Fig.\ref{fig:Fcrit_vs_singleShot}.b, we observe single shot helicity-independent AOS for pulses as long as $\Delta t=15$ ps pulse in the Gd$_{27}$FeCo sample. For $\Delta t > 15$ ps $F_C$ exceeds the multidomain critical fluence. The result is then a fully demagnetized pattern (pic \#4 in Fig.\ref{fig:Fcrit_vs_singleShot}.b) and no AOS is ever observed for these pulse durations. This is in sharp contrast with the $\Delta t=55$ fs multidomain state (pic \#3 in Fig.\ref{fig:Fcrit_vs_singleShot}.a) where the outer part of the Gaussian laser beam does induce AOS.

\begin{figure}
\includegraphics[trim=0cm 0cm 0cm 0cm, clip=true, width=1\columnwidth]{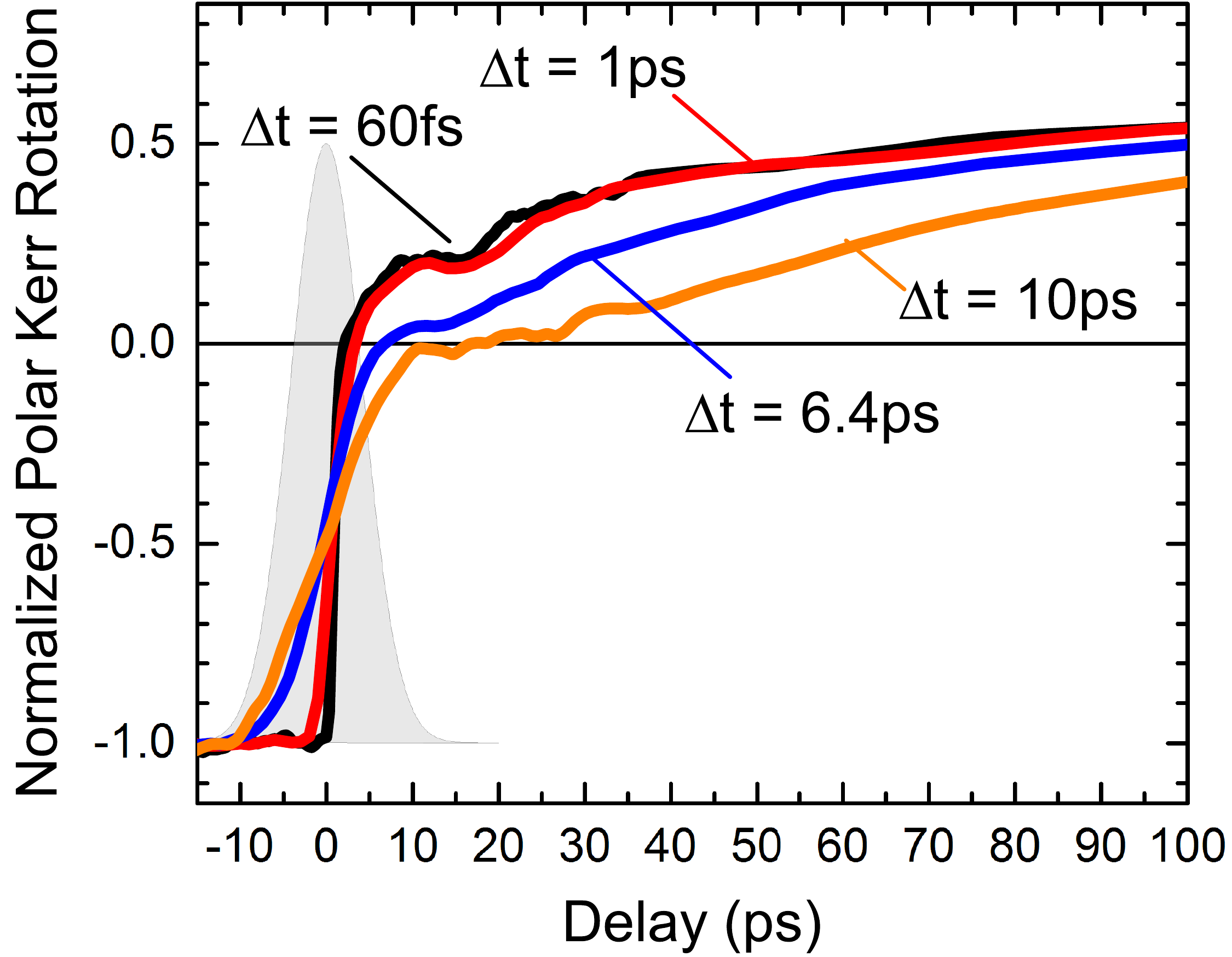}
\caption{\label{fig:pumpprobePulse}: In solid lines, the evolution of the magnetization of Gd$_{27}$FeCo after a linearly polarized pump pulse, at room temperature, for $\Delta t = 60$ fs , $1$ ps, $6.4$ ps, $10$ ps and $F_C \sim 0.8$ mJ/cm$^2$, $0.9$ mJ/cm$^2$, $1.0$ mJ/cm$^2$, $1.6$ mJ/cm$^2$ respectively. A $10$ ps pump intensity profile is depicted in light grey. The probe duration was kept equal to the pump duration, which results in a loss of resolution and a smoothing of the long pulse duration curves. The switching time (crossing of $0$) increases with the pump duration, and always happens after all the energy has been deposited on the film. Note that zero time delay was readjusted since tuning the compressor introduces small changes ($\sim2$ mm) in the pump and probe paths. Zero time delay was set by assuming that the maximum slope of the demagnetization corresponds to the peak of the pump pulse.}
\end{figure}

We performed time resolved pump probe experiments on Gd$_{27}$FeCo at various pulse durations in order to see how the switching dynamics are affected by the electron's heating rate. This is shown in Fig.~\ref{fig:pumpprobePulse}. The probe pulse duration was kept equal to the pump, which results in a loss of time-resolution and smoothing of the data for longer pulse durations. Due to a decrease of the AOS fluence window at long pulse durations, the probe was tightly focused through a 50x objective onto  $\sim 2$ $\mu$m spot at the center of the pump spot. A constant $\sim200$ Oe magnetic field was applied to reset the magnetic state of the film between pulses. As the pump duration increases, the switching time (crossing of $0$ on the $y$ axis) increases from $\sim 2 $ ps up to $\sim 13 $ ps for $\Delta t =55$ fs and $10$ ps respectively. The switching happens in all cases after all of the energy of the optical pulse is deposited in the film. This result shows that using $10$ ps optical pulses, we can still perform a rather fast switching of the magnetization, which releases the constraint on using femtosecond lasers for the study of AOS and for applications.

\begin{figure}
\includegraphics[trim=0cm 0cm 0cm 0cm, clip=true, width=1\columnwidth]{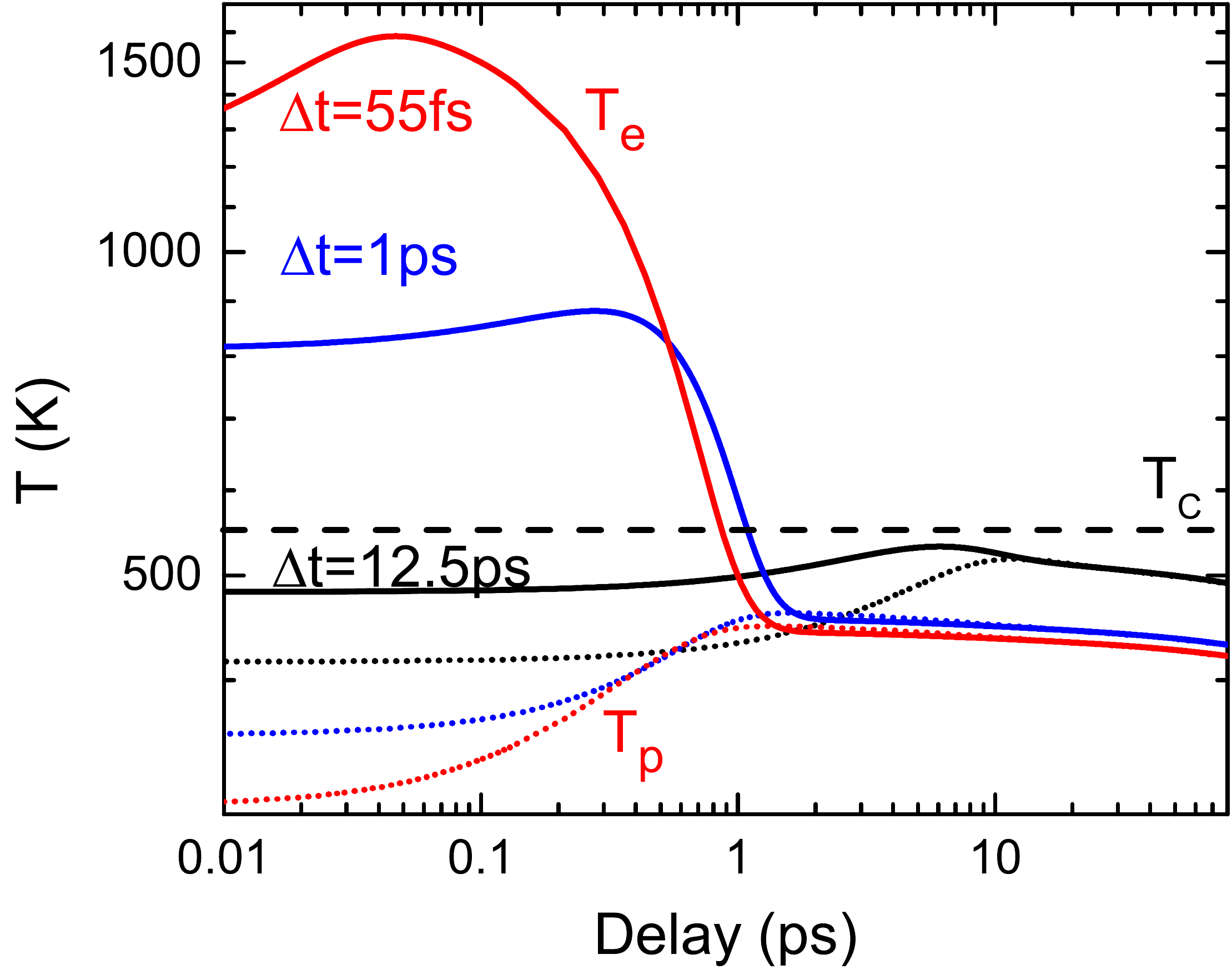}
\caption{\label{fig:2TM}: Calculation of electronic (solid lines) and lattice (dotted lines) temperatures after a $\Delta t=55$ fs $F_C=0.8$ mJ/cm$^2$ pulse (red), a $\Delta t=1$ ps $F_C=0.9$ mJ/cm$^2$ (blue) and a $\Delta t=12.5$ ps $F_C=1.35$ mJ/cm$^2$ pulse (black) according to the three temperature model (see text). The dashed line indicates $T_C$. For $\Delta t=12.5$ pulses, $T_e$ gets very close to $T_C$. Whether $T_e$ needs to reach $T_C$ or not is unclear due to the uncertainties ($\sim 20$\%) of the critical fluences $F_C$.}
\end{figure}

The transient temperature response of the electrons and phonons during AOS with $\Delta t=55$ fs, $\Delta t=1$ ps, and $\Delta t=12.5$ ps pulses at fluences equal to $F_C$ are shown in Fig.~\ref{fig:2TM}.  We calculated the temperature responses using the three temperature model~\cite{Kimling2014}. 
We fixed the electron heat capacity $C_e = \gamma T_e$, with $\gamma = 300$ J/(m$^3$K$^2$) based on first principle band structure calculations of amorphous GdFe$_2$~\cite{Tanaka1992}. The lattice heat capacity is set to $2.3$ J/(m$^3K$) , a weighted average of the lattice heat capacity of Ta and GdFe$_2$~\cite{Hellman1998}.  The spin heat capacity in our model as a function of temperature was fixed by subtracting the electronic and lattice heat capacities from the total heat capacity of GdFe$_2$~\cite{Hellman1998}.  The electron-spin coupling constant was fixed to $10^{17}$ W/(m$^3$K) and the electron-phonon coupling constant was set to $6 \times 10^{17}$ W/(m$^3$K). These two values were set based on separate thermal transport measurements of Au/GdFeCo metallic bilayers that we have made and will report elsewhere~\cite{ephCoupling}.

We do not consider the spin temperature in our three temperature model calculation to be a valid descriptor of the thermodynamic state of the spin system. The transient magnetic states that occur following laser irradiation do not occur in the equilibrium phase diagram of GdFeCo, and therefore cannot be described with an effective temperature. Therefore, the sole purpose of the spin temperature in our model is to account for the impact of energy transfer between the electrons and magnetic sublattices on the transient temperature response of the electrons. This channel for energy exchange needs to be considered, especially when close to $T_C$ where the magnetic heat capacity is as large as $\sim40$\% of the total heat capacity.


The small increase in $F_C$ as $\Delta t$ increases implies that the peak electron temperature of the system is not particularly important for the helicity-independent AOS. In fact, as shown in Fig.~\ref{fig:2TM}, for $\Delta t=12.5$ ps pulses, $T_e$ will only be heated to $\sim 530$ K. We are not able to exactly determine whether $T_e$ needs to reach $T_C$ or not, due to the uncertainty ($\sim 20$\%) in the critical fluence. Despite this open question, our result raises questions on the proposed scenario where very high electron temperatures ($1000-2000$ K) are necessary for AOS~\cite{Radu2011, Ostler2012a, Mentink2012, Barker2013, Wienholdt2013,Atxitia2015,Kalashnikova2016,Schellekens2013b}.




We posit that helicity independent switching is a three step process, where there is no need for high electron temperatures. First, after optical absorption, the energy per Fe spin degree of freedom becomes slightly higher than the energy per Gd spin degree of freedom (i.e. the Fe is hotter), as proposed by Wienholdt et al.~\cite{Wienholdt2013}. Second, the Fe and Gd spins exchange energy and angular momentum on a time-scale faster than the time scale of angular momentum dissipation into the lattice. This corresponds to a semi-adiabatic process and the dynamics of the system are thus governed by the principle of maximization of entropy as described by, 

\begin{eqnarray}
\label{eq:Thermo}
(2J_{FG}S_{F}-2J_{FF}S_{G}+J_{FG}S_{F}-J_{GG}S_{G})dS_{F}>0
\end{eqnarray}

where the left side of the equation corresponds to the change in internal energy of the system. $J$ is the exchange constant and $S$ is the total spin angular momentum of sublattices Fe ($F$) or Gd ($G$). In GdFeCo, $J_{FF}$ and $J_{GG}$ are negative,  $J_{FG}$ is positive, so that $S_{F}$ and $S_{G}$ have initially opposite signs. Conservation of angular momentum ($dS_{F}=-dS_{G}$) is implied.

To fulfill Eq.~\ref{eq:Thermo} we find that $\mid S_{F}\mid$ and $\mid S_{G}\mid$ must decrease, meaning demagnetization of the sublattices will occur. If the Fe sublattice is initially hotter,  the Fe will reach full demagnetization first. With the Fe fully demagnetized ($S_{F}=0$) Eq.~\ref{eq:Thermo} implies the switch and growth of the Fe sublattice parallel to the Gd spins, leading to a transient equilibrium ferromagnetic state~\cite{Wienholdt2013}. In other words, on time-scales over which angular momentum is conserved, the temporary equilibrium state will be ferromagnetic because entropy is maximized with ferromagnetically aligned Gd and Fe spins

In the third and last step, the Gd switches in order to be antiferromagnetically aligned with the now hot and dominating Fe lattice~\cite{Kirilyuk2013} and both sublattices remagnetize as they cool down. Remagnetization occurs on much longer time-scales than demagnetization, so spin angular momentum is not conserved anymore.

In the proposed three step scenario, the magnetization can switch sign without ever reaching the Curie temperature. There are two requirements. First, the Fe spin system must be preferentially heated with respect to the Gd spins~\cite{Wienholdt2013}. Second, exchange of energy between sublatices should happen faster than the timescales of dissipation of angular momentum into the lattice~\cite{Chimata2015}. Moreover, the lattice temperature should remain below $T_C$ at all times, otherwise resulting in a multidomain final state.


The difference in maximum pulse duration between samples Gd$_{24}$FeCo and Gd$_{27}$FeCo shown in  Fig.~\ref{fig:Fcrit_vs_singleShot}.b resides most probably in the differences in energy transfer rates between the Fe and Gd spin sublattices, which depend on the composition~\cite{Barker2013}. Experimentally examining the relationship between alloy composition and energy transfer rates will be the subject of a future work.



In summary, we carefully extracted critical fluences for AOS in GdFeCo as a function of the initial temperature of the sample and pulse duration by single-shot and stroboscopic measurements. We confirm that lattice heating is not the main mechanism for AOS. We then showed that AOS is possible for pump laser pulse duration up to $15$ ps. We performed pump-probe experiments as a function of the pulse duration and showed that the switching time increases as the pump duration increases, with $10$ ps pulses resulting in switching times of $\sim 13$ ps. We estimated the temperature rise for electrons and the lattice via the three temperature model and showed that the peak electron temperature is not a major parameter for AOS as it varies from $\sim 1600$ K for $\Delta t= 55$ fs to $\sim T_C$ for a $\Delta t=12.5$ ps. AOS with $15$ ps pulses challenges previous models for helicity-independent AOS where high electron temperatures are assumed responsible. Finally we suggest a three step thermodynamical model of the switching based on the preferential heating of Fe spins compared with Gd spins, and on the fast energy exchange between the sublattices.


{\it - Acknowledgements:}
This work was primarily supported by the Director, Office of Science, Office of Basic Energy Sciences, Materials Sciences and Engineering Division, of the U.S. Department of Energy under Contract No. DE-AC02-05-CH11231 within the Nonequilibrium Magnetic Materials Program (MSMAG). Sample fabrication was supported by C-SPIN: one of the six SRC STARnet Centers, sponsored by MARCO and DARPA. We also acknowledge the National Science Foundation Center for Energy Efficient Electronics Science for the providing most of the experimental equipment and partially supporting operation of the experiments.

\bibliography{library,library_GdFeCo}

\end{document}